\documentclass{sig-alternate}

\usepackage{subfigure}
\usepackage{rotating}
\usepackage{paralist}
\usepackage{url}
\usepackage{lscape}
\usepackage{multirow}
\usepackage{amsmath}
\usepackage{mathtools}
\usepackage{tabularx}

\begin{document}

\tolerance=2000
\clubpenalty = 10000
\widowpenalty = 10000

\conferenceinfo{PROMISE}{'13, October 9, 2013, Baltimore, USA}
\CopyrightYear{2013} 
\crdata{978-1-4503-2016-0/13/10}  

\title{Are Comprehensive Quality Models Necessary\\ for Evaluating Software Quality?}

\numberofauthors{2}

\author{
Klaus Lochmann
\\[6pt]
 \affaddr{Technische Universit\"at M\"unchen}\\
       \affaddr{Software \& Systems Engineering}\\
       \affaddr{Munich, Germany}\\
       \email{lochmann@in.tum.de}
\alignauthor       
Jasmin Ramadani, Stefan Wagner
\\[6pt]
 \affaddr{University of Stuttgart}\\
       \affaddr{Institute of Software Technology}\\
       \affaddr{Stuttgart, Germany}\\
       \email{jasmin.ramadani, stefan.wagner@informatik.uni-stuttgart.de}
}
\date{20 March 2013}

\maketitle
\begin{abstract}
The concept of software quality is very complex and has
many facets. Reflecting all these facets and at the same time
measuring everything related to these facets results in comprehensive but large
quality models and extensive measurements. 
In contrast, there are also many
smaller, focused quality models claiming to evaluate quality with few measures.

We investigate if and to what extent it is possible
to build a focused quality model with similar evaluation
results as a comprehensive quality model but with far less measures 
needed to be collected and, hence, reduced effort.
We make quality evaluations with the comprehensive Quamoco base quality
model and build focused quality models based on the same set of measures
and data from over 2,000 open source systems. We analyse the ability of
the focused model to predict the results of the Quamoco model by
comparing them with a random predictor as a baseline. We
calculate the standardised accuracy measure SA and effect sizes.

We found that for the Quamoco model and its 378 automatically collected measures,
we can build a focused model with only 10 measures but an accuracy of 61\,\%
and a medium to high effect size. We conclude that we can build focused 
quality models to get an impression of a system's quality similar to comprehensive 
models. However, when including manually collected measures, the accuracy of the 
models stayed below 50\,\%. Hence, manual measures seem to have a high impact and 
should therefore not be ignored in a focused model.
\end{abstract}

\category{D.2.9}{Software Engineering}{Management}[Software Quality Assurance (SQA)]

\terms{Experimentation}

\keywords{Software quality, quality model, quality evaluation}

\section{Introduction}

Software quality is a core concept of software engineering as we aim to
build software systems in time, on budget, with suitable functionality and
high quality. Almost any activity and method in software engineering
influences the quality of software. To describe, evaluate and predict
quality, we use quality models. They help us to understand what we need
to pay attention to when building the system and also what to measure
when we want to know the quality level of the system.


The fact that software quality is very complex and has many facets suggests
that we also need comprehensive quality models to cover all these facets.
Recent initiatives such as CISQ~\cite{Soley.2010}, Squale~\cite{MordalManet.2009}
and Quamoco~\cite{Wagner.2012} propose to some degree similar solutions
in the form of large, operationalised quality models to support detailed measurement
as well as the breadth of software quality.
As a result, the quality models 
demand to collect a lot of data for various measures, some of them manually,
with high effort or expensive tools. 
This, in turn, leads to comprehensive but also
very large quality models. For example, the Quamoco base model contains 194
factors with an influence on quality measured by 526 measures for different
programming languages.

\subsection{Problem Statement}
This comprehensiveness and level of detail of the comprehensive quality models have 
several advantages: They can help
developers and quality engineers to better understand the meaning of measures
and the relations between different factors with an impact on quality. Furthermore,
they support the specification of detailed, testable quality requirements. Also for a
comprehensible and precise quality evaluation, it is helpful to have all these details.

In contrast, there are many focused quality models, usually based on a small set
of measures and regression or machine learning, to evaluate one or many quality
aspects. They can be useful in practical settings where we prefer a less comprehensive and precise
evaluation but at lower cost or effort. The various measures in the comprehensive models can come
from diverse tools at differing costs and also for manual inspections with high effort. In software
quality, however, a small, single defect can have a huge impact. Not including detectors
for such defects can hence be a risk for the validity of the quality evaluation.
So far, it is not clear what effects this trade-off between comprehensiveness
and low application effort has. Even between the focused quality models,
there are significant differences in the evaluation results~\cite{Lincke.2010}.

\subsection{Research Objective}
Our overall aim is to make the application of quality models and their
integration into the development process as simple and effortless as
possible. To give immediate feedback to developers when quality problems
have been introduced, we propose continuous quality control~\cite{Deissenboeck.2008}. 
That ideally includes a continuous quality evaluation as well. Hence,
our goal is to make the evaluation using a quality model fast and with low effort.
In this paper, we concentrate on investigating the trade-off in using large,
\emph{comprehensive quality models} or \emph{focused quality models} that
use a limited set of measures for quality evaluation.

\subsection{Contribution}
Different quality models predict different measures as quality evaluation and,
hence, are difficult to compare. Instead, we use an existing, comprehensive quality model
and build focused quality models using various machine learning approaches based
on the same set of measures that the comprehensive quality model uses. This way, we can
analyse the accuracy of the prediction models.
We use the maintainability part of the Quamoco base model as example for
a comprehensive quality model, because its evaluations have a good 
correspondence to expert opinion~\cite{Wagner.2012},
and established approaches, such as random forrest and regression, for 
building the focused models.
In addition, we investigate the difference
if manually collected measures are included. This is important, because
those measures incur the highest costs to collect. We perform all the comparisons
using an established approach: random partitions for
model building and testing, comparison to a random predictor as baseline, the
standardised accuracy measure SA~\cite{Shepperd.2012} and effect sizes~\cite{Cohen.1992}.

Our results show that it is possible to build a focused quality model
based on a predictor: The best of the tested predictors relies only on
10~measures, compared to several hundred ones in the full model, but 
its accuracy is 61\,\% with an effect size of 0.70.
Furthermore, our results show that the expert-based measures cannot be 
neglected, as the focused model built without expert-based measures 
works considerably worse for predicting evaluations which include expert-based measures.



\section{Quamoco Quality Model}
\label{sec:quamoco}

In the research project \emph{Quamoco}, we developed a quality model for defining 
and evaluating quality~\cite{Wagner.2012}. The aim was that the resulting quality 
model should bridge the gap between abstract quality characteristics and concrete 
measures. Thus, the quality model defines \emph{factors}, as an abstraction level
between abstract quality characteristics and quantifiable measures. Such a comprehensive 
quality model support developers and quality engineers in different tasks. For instance, 
a Quamoco quality model can be used in requirements engineering as a checklist, so 
that one does not forget important quality characteristics and then add concrete 
factors and measurements so that the quality requirement is testable. The task a 
quality model can support best, however, is the evaluation of the quality of a software 
product.

We built the Quamoco base quality model as a collection of widely applicable
quality factors and measures for the programming languages Java and C\#. We describe
the concepts and structure of the base model, then its contents and, finally, the quality
evaluation method based on the quality model.

\subsection{Concepts and Structure}

\begin{figure}[!ht]
	\vspace{0.5em}
	\centering
	\includegraphics[width=\columnwidth]{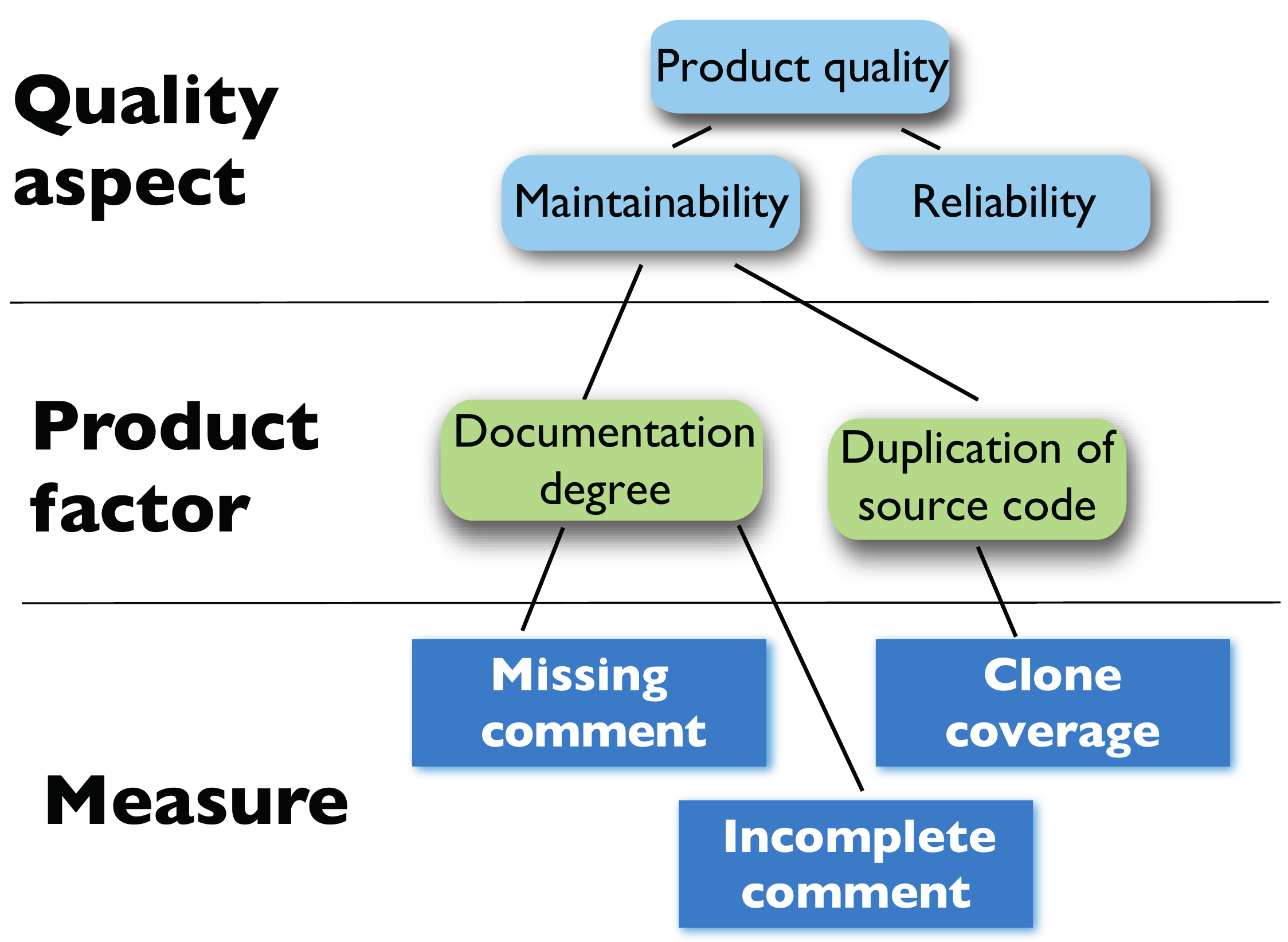}
	\caption{Schematic Structure of the Quamoco Base Quality Model}
	\vspace{0.5em}
	\label{fig:quamoco-model}
\end{figure}

We started the Quamoco project, because we perceived a gap between abstract
quality characteristics and concrete measurements~\cite{Wagner.2012}. We found
it very hard to assign very detailed measures to the very broad quality characteristics, which
are hard to further break down, and aggregate them to a sensible quality evaluation result. 
Therefore, we kept the abstract level, which we called the \emph{quality aspect} level, and the concrete level
which we called the \emph{measure} level. As shown in Figure~\ref{fig:quamoco-model},
we added an additional level, on which we describe factors of the product (properties
of entities) that have an impact on the quality aspects and are measured by the
measures. This is, for example, similar to the quality-carrying properties of Dromey~\cite{Dromey.1995}.

Quamoco quality models have more elements and details as shown in 
Figure~\ref{fig:quamoco-model}. For instance, all the quality aspects, product factors
and measures have IDs and descriptions. The impacts between product factors and
quality aspects can be positive or negative and have a justification. There is additional
information on how measurement results can be collected. We concentrate, however,
on the information necessary for the following.

In principle, one can use any quality aspect hierarchy but to be consistent with others,
we included the ISO/IEC~25010 product quality hierarchy in the Quamoco base quality
model. The product factor level contains then various properties of the product, mostly
concentrating on the source code. The \emph{documentation degree} of 
Figure~\ref{fig:quamoco-model}, for instance, has the following description:
\begin{quote}
	A source code's documentation degree is high if it is commented as needed.
\end{quote}
This product factor is not directly measurable, and thus we operationalize it
by measures, which check whether comments are missing or incomplete. For missing 
comments, we have, for example, the tool ConQAT\footnote{\url{http://www.conqat.org}}
which gives us locations in Java source code where there should be a JavaDoc comment
but it is not there. Missing comments are only part of the evaluation of documentation degree.
It is also important what the comments contain. Therefore, there is also the manual
measure \emph{incomplete comment} which gives the proportion of incomplete
comments in the source code. Together, the two measures can give a good impression
of the degree of documentation of the source code.

To aid the structuring of product factors in a hierarchy and the clear definition of
product factors, we associate an \emph{entity} with each product factor. An entity
is a part of the software product, e.g.~the whole source code, sub-routines or statements.
For \emph{documentation degree} the entity is the \emph{comment} in the source code.

We have done intensive empirical validations of the Quamoco approach and especially
the Quamoco base model~\cite{Wagner.2012, Lochmann.2011}. We found that quality evaluations using
the base model led to results that corresponded well with expert opinion. Especially the
maintainability part of the base model seems to give very accurate evaluations. Therefore,
we assume that the base model gives valid estimates of the quality level of a software
system.

\subsection{Contents of the Quamoco Base Model}
\label{sec:qm}

The base model's main objective is to describe software 
quality in a way that a wide range of software products can be evaluated without
requiring large effort upfront. To reach this goal, the Quamoco project partners
conducted various workshops to collaboratively transfer their 
knowledge and experience into the Quamoco model structure.

The resulting quality model represents our consolidated view on
the quality of software source code and is, in principle, applicable to any
kind of software. It details quality down to particular
analysis tools as instruments for the evaluation of Java and C\# systems,
and, hence, enables comprehensive, tool-supported quality evaluation 
without requiring large adaptation or configuration effort. 

In total, the base model comprises more than 1,500 model elements. Out of these,
there are 201 product factors with an impact to quality aspects measured by 526
measures for both Java and C\#. This reflects that a product factor can be measured by
more than one measure and it can also be relevant for several programming languages.
The measures contain 8 manual ones and 518 that are provided by one of 12 
different tools. The tools most relied upon are FindBugs (Java, 361 rules 
modelled) and Gendarme (C\#, 146 rules). Other tools integrated into our model 
include PMD and several clone detection, size and comment analyses, which are 
part of the quality assessment framework ConQAT. 

To illustrate the contents of the base model, we will describe two examples
of product factors together with their measures and impacts.
%
As described above, the largest fraction of measures refers to static
code analysis tools. One example is the FindBugs rule   
FE\_TEST\_IF\_EQUAL\_TO\_\-NOT\_A\_NUMBER, which scans Java code
for equality checks of floating point values with the \emph{Double.NaN} constant.
The Java language semantics defines that nothing ever equals \emph{NaN},
not even \emph{NaN} itself, so that \emph{(x == Double.NaN)} is always false. 
To check whether a value is not a number, the programmer has to call \emph{Double.isNaN(x)}.
This rule is an instrument for the \emph{doomed test for equality to NaN}
measure, which measures the factor \emph{general expression applicability}
for \emph{comparison expressions}, along with several other measures.
This factor in turn impacts \emph{functional correctness}, because the developer
intended to check a number for \emph{NaN}, but the actual code does not.
It furthermore impacts \emph{analysability} because to understand the intention of
this construct demands additional effort.

Rule-based code analysis tools cannot detect every kind of quality problem.
Therefore, the base model also contains product factors based on 
established research results, metrics, and best practices. For instance, identifiers have
been found to be essential for the understandability of source
code. Whether identifiers are used in a concise and consistent
manner can only partly be assessed automatically~\cite{Deissenboeck.2006}.
Therefore, the factor \emph{conformity to naming convention} for source
code identifiers contains both automatic checks performed by
several tools and manual measures for assessing whether identifiers
are used in a consistent and meaningful way.


\subsection{Evaluation Method and Tooling}

We evaluate the quality of a software product using the quality model by adding
\emph{evaluation specifications} to product factors and quality aspects. For both,
we evaluate on the scale $[0,1]$ expressing the degree to which 
the factor is present in the system. An evaluation of $0$ means the factor is
not present in the system, while $1$ means the factor is fully present in the system.
The evaluations of product factors are mappings from the values of its associated
measures to the scale $[0,1]$. We define a minimum threshold and a maximum 
threshold; below the minimum threshold 0 is assigned and above the maximum threshold
1 is assigned. Between these thresholds, the evaluation changes linearly.
The quality aspects evaluate simply by aggregating the product factors that have an
impact on them. The aggregation is done by weighted sums similar to~\cite{Alves.2010}.

Our goal was to use very simple evaluations and aggregations so that it is
easily comprehensible for practitioners. The challenge lay in defining reasonable
thresholds for the evaluations. We calibrated these evaluations by measuring a 
large number of open source systems to find out what are ``normal'' values~\cite{future-wosq-2012}. These
defined the thresholds, we also reviewed and corrected based on expert
opinion. For example, the measure \emph{clone coverage} denotes the probability
that for a randomly chosen statement in the source code, there exists a copy. We calculated
the quartiles of the data from 125 Java systems, removed the outliers and used the
minimum (0.0) and maximum (0.57) as thresholds. 

The scale from 0 to 1 is very general and, hence, suitable to integrate various
measures and factors. To help the quality engineer in interpreting the results for quality 
aspects, we added an additional step: the transformation into a grade
scale. Being a German research project, we chose the German school grades as
our grade scale. It goes from 1 to 6, 1 being the best and 6 the worst grade. The analogy
we used was a dictation at school. If you have more than 10\,\% of your text wrong, you
will get a 6. Hence, 0--0.90 gives a 6, 0.90--0.92 a 5 and so on. This turned out to be
well comprehensible for practitioners~\cite{Wagner.2012}.

For creating quality models and conducting automated quality
evaluations, we developed a tool chain~\cite{Deissenboeck.2011}. 
It consists of a graphical quality model editor used for quality 
modelling and the quality assessment toolkit ConQAT\footnote{\url{http://www.conqat.org/}} 
used for analysing artefacts and data. ConQAT and the Quamoco
quality model editor are open source software 
licensed under the Apache~2.0 license.


ConQAT is a toolkit for the creation of quality dashboards
that is configurable with a graphical domain-specific language. 
It provides diverse quality analyses out of the box and integrates
with code analysis tools like Findbugs, PMD, or Gendarme. 
The results of the quality analyses are visualised in HTML dashboards.
For conducting quality evaluations based on the quality model,
the quality model editor allows to automatically generate a
ConQAT configuration that contains all the required analyses
to evaluate the software product according to the 
specifications in the quality model. 

\section{Research Strategy}
\label{sec:approach}

We detail in the following how we conducted our experiment to compare
the evaluations of the comprehensive Quamoco base model with a set of focused
quality models built by machine learning.

\subsection{Research questions}
Our question overall is: \emph{Are comprehensive quality models necessary for
evaluating software quality?} Hence, we want to know if we can come to
similar evaluation results using focused models, in our case built from
existing data using machine learning approaches. In order to apply machine
learning in a meaningful way, we need a large enough set of existing data.
Thus, we will use the part of the Quamoco base model focusing on maintainability, 
because this part is the most comprehensive and well-elaborated one. It
includes 378 maintainability-related measures and evaluates and aggregates 
them to a single grade for maintainability.

To further structure our study, we use two research questions reflecting
the general performance of the focused models as well as the effects
of expert-based measures:
\begin{itemize}
	\item RQ~1: What is the performance of focused quality models built using machine learning algorithms?
\end{itemize}
The large number of measures in the Quamoco base model reflects the diversity of influences in the code onto maintenance. Under certain circumstances, however, we would like to be able to get an evaluation that is less accurate but needs less measures. For example, tools might not be available or analyses run too long for an hourly analysis. Therefore, we investigate to what extent it is possible to build a predictor, which calculates an evaluation result using a small number of measures.
\begin{itemize}
	\item RQ~2: What is the performance of the focused quality models including additional expert-based measures?
\end{itemize}
An especially interesting use case for a focused quality model would be to avoid manual, expert-based
measures. They are usually collected during manual reviews, which means they are elaborate to collect and
cannot be frequently measured. We build the predictors for RQ~1 using a large number of systems, for which 
only fully automated measures are available, however. For only 15 systems we have expert-based measures
which are part of the Quamoco base quality model. In this research question, we focus on how the predictors 
perform for these 15 systems.

\subsection{Predictor models}
As focused quality models we use predictor models as known from machine learning, which predict the quality
evaluation results of the comprehensive quality model. We use the following common machine learning algorithms to 
create the predictor models:
\begin{itemize}
	\item Linear Regression: We use multiple linear regression, which is modelling the distribution 
				of a \emph{dependent variable}, using one or more \emph{independent variables}. Linear 
				regression models the relation between the dependent variable and the independent variables 
				by fitting a linear equation to observed data. In our case, the independent variables are
				the measures of the Quamoco quality model, and the dependent variable is the evaluation 
				result for maintainability. We do not use data transformation, as the inspection of the 
				residual plots suggested a random distribution of the residuals and thus a good fit of the linear model.
	\item Classification Trees: We use the prediction method of classification trees according to~\cite{Breiman.1984}. 
				This machine learning approach builds a tree in which the leafs represent the values of the
				dependent variable, the nodes represent branches based on the independent variables. Such a
				tree is constructed by recursively partitioning the data and deriving a branching condition
				for the resulting node. This approach is only applicable to categorical data, as in our case 
				the school grades.
	\item Random Forests: Random Forests~\cite{Breiman.2001} are defined as a combination of tree 
				predictors, whereby each tree is constructed based on a random selection of samples from 
				the independent variables and Breiman's idea called bagging which reduces variance and 
				helps to avoid overfitting. Using Random Forests we grow many classification trees where 
				to classify an object the input has to be sent through all of the generated trees in the 
				forest. Each of the trees performs classification and ``votes'' for the most popular class. 
				At the end the forest is choosing the classification with the most votes by all of the trees in the forest.
\end{itemize}

Since not all predictors allow us to influence the number of variables they use, we apply a 
forward selection approach~\cite{Freund.1998} for most predictors to incrementally increase the 
number of used variables. The forward selection is a simple and often used data-driven model 
building approach. In this approach, we start with an empty model containing no independent variables.
The independent variables are added one at a time, testing all not yet included variables in each step 
and adding the most significant variable. The procedure is repeated until there are no more improvements 
in the model by adding another independent variable.

Concerning these machine learning algorithms, we use the following predictor models to 
investigate if they are useful for creating a focused quality model:
\begin{itemize}
	\item \emph{Random Guessing}: Instead of doing an actual predition based on independent
				variables, the random guessing approach returns a random value as predicted value. 
				The distribution of the returned values is the same as that of the dependent variable. 
				We use the random guessing approach as baseline to see how much the other	predictors 
				improve over it. 
	\item \emph{$5\,\%$-Quantile of the Random Guesses}~($5\,\%$ of the best guesses): This predictor
				looks at the 5\,\% of the best guesses. It means that any value of accuracy that is better 
				than this threshold has a chance of less than 1 in 20 to be a random occurrence. 
	\item \emph{Linear Regression Model with Forward Selection}: This predictor combines the
				linear regression and the forward selection algorithm described previously.
	\item \emph{Linear Regression Model with Backward Elimination}: This predictor model applies
				the inversion of the forward selection to the linear regression. It starts with the 
				maximum model consisting of all independent variables. The deletion of each variable 
				is tested at each step and the variable that has the smallest significance is deleted. 
	\item \emph{Linear Regression Model with Bidirectional Elimination}: This predictor applies
				a combination of forward selection and backward elimination to the linear regression. 
				At at each step, all independent variables are tested for either their inclusion or 
				exclusion. 
	\item \emph{Classification Tree with Different Complexity Parameters}: When constructing 
				classification trees it is possible to influence the number of used independent 
				variables by predefining a required fit (denoted complexity parameter) of the 
				independent variables with the dependent variable. The tree construction algorithm 
				then selects the needed number of independent variables to achieve the required
				fit. We construct a series of classification trees with different complexity parameters
				and thus a different number of independent variables. 
	\item \emph{Classification Tree with Forward Selection}: Instead of controlling the number of
				used independend variables by the complexity parameter, for this predictor we apply the 
				forward selection approach to classification trees. 
	\item \emph{Random Forest with Forward Selection}: Since the algorithm for constructing random
				forest predictors does not allows any control over the number of independent variables used,
				here we apply the forward selection approach to random forests.
\end{itemize}

\subsection{Model Comparison}
\label{sec:metrics}

We need statistics for comparing the performance of predictors. A commonly used statistics for
assessing the performance of a predictor is to calculate the magnitude of relative error (MMRE).
Shepperd and MacDonell~\cite{Shepperd.2012} state that ``biased accuracy statistics such as MMRE are
deprecated.'' They introduce a new validation framework to assess and compare prediction systems. We
will use their framework and introduce the main statistics proposed by them in the following.

For measuring the performance of a predictor, we use the statistic \emph{mean absolute residual}~(MAR), 
defined as follows:
\begin{equation}
	MAR=\frac{\sum^n_1{|}\left({y}_{i}-{\hat{y}}_{i} \right)|}{n}
\end{equation}
MAR is the mean of the absolute error values for the observations providing information with regard to the 
prediction model fit. It has the advantage of being an unbiased statistic, because it is not based on ratios.

To interpret the MAR statistic, Shepperd and MacDonnell introduce the idea of comparing a given 
predictor with a baseline predictor. As baseline predictor we take a simple approach of using the random 
guessing predictor. This results in the \emph{standardised accuracy measure}~(SA), which is the MAR relative 
to random guessing. For a suggested prediction technique $p_i$, where \emph{i} denotes a prediction technique 
used, the standard accuracy measure is defined as the following:
\begin{equation}
	{SA}_{p_i}=1-\frac{MAR_{p_i}}{\overline{MAR}_{p_0}} 
\end{equation}
where $\overline{MAR}_{p_0}$ is the mean value of a large number of random guessing runs. It is defined as 
prediction of $\hat{{y}}_{i} $ for the target case $t$ by random sampling over all the remaining $n-1$ cases 
and takes $\hat{{y}}_{t}={y}_{r}$ where $r$ is randomly chosen from $1\dots n \wedge r \neq t$. After many runs, 
the $\overline{MAR}_{p_0}$ converges on using the sample mean. The interpretation of SA is that the ratio shows 
how much ${p}_{i}$ is better than random guessing ${p}_{0}$. A value close to zero or a negative value is considered 
bad for the predictor model. A low MAR and a resulting SA close to 100\,\% would mean that the predictor model is almost 
perfect.
								
In addition to the SA statistic, we use the effect size statistic to investigate how big the difference is really in 
relation to the overall variation in the random guesses. It is defined by Glass's $\Delta$~\cite{Shepperd.2012} as follows:
\begin{equation}
	\Delta =\frac{{MAR}_{{p}_{i}}-{\overline{MAR}}_{{p}_{0}}}{{s}_{{p}_{0}}}
\end{equation}
where ${s}_{{p}_{0}}$ is the sample standard deviation of random guessing. The Glass's $\Delta$ is a biased estimator 
for small sized samples or where there is a large difference in the size of the samples. It standardises the difference 
between two different prediction models and it gives context to the difference of amount of variations in the measures. 
The effect size as suggested by Cohen~\cite{Cohen.1992} can be categorised as small ($\approx$0.2), medium ($\approx$0.5) 
or large ($\approx$0.8).

\subsection{Finding a Model with a Low Number of Measures}
Our goal is not to just find a predictor which performs well in predicting the quality evaluations of the
comprehensive quality model, but to find a predictor which does so using a low number of measures. Thus, 
for each type of predictor, we construct a series of it, each using a different number of measures. For
analysis and discussion, we then select two predictors for each type: First, we take a look at the 
predictor having the best result in terms of the SA statistic. Second, we select the predictor using the
minimal number of measures, so that its MAR is not more than 10\,\% less than the MAR of the best-performing
predictor. Calculating percentages of the MAR statistic is valid since it is an absolute statistic.

\subsection{Study Objects}
\label{sec:study_objects}

For conducting automated quality evaluation, we need software products to be evaluated.
We used the repository SDS~\cite{SushilBajracharya.2009,JoelOssher.2009}, 
containing about 18,000 open-source Java projects. These projects have mostly been retrieved 
from open source databases such as \emph{Sourceforge} through a web-crawling approach. In essence, 
this repository contains mirrors of the version control repositories of the before-mentioned
databases. The SDS repository only contains the source code, not the binaries. For the quality 
evaluation, however, binaries compiled with the debug-option of the Java compiler are 
needed. We compiled all projects in a batch approach, because the effort to manually 
configure and compile them is prohibitive. The compilation of all 18.000~projects took 
about 30~hours, executed in parallel on 12~personal computers. Of all available projects 
about 6.000 compiled successfully. Others could not be compiled because of missing 
external libraries or because of code needed to be generated during the build process. 
We used all systems of the SDS repository larger than $5.000$~LoC as a benchmarking base,
resulting in 2041 systems. We excluded smaller systems, because many open source repositories 
contain software projects initiated by single persons without finishing them; these projects 
then remain in the repository without ever being used~\cite{Beecher.2009,Rainer.2005}. Of the
2041 systems that compiled, the quality analysis tool run successfully on 1994 systems. Thus,
these 1994 systems are used as study objects.
The distribution of sizes shows that half of the systems are smaller than 11\,kLoC 
and 90\,\% of the systems are smaller than 50\,kLoC. Only 54 systems are larger than
100\,kLoC, with the largest system having 477\,kLoC.

Regarding RQ~2 we have 15 systems available, for which the expert-based measures of
the quality model have been collected. Besides our own tool (ConQAT), we selected 
the most downloaded SourceForge\footnote{\url{http://sourceforge.net/}, accessed in December 2010.} 
projects written in Java. Of these projects we included only those, where it was 
possible to compile the source code on our own with reasonable effort (less than 
one hour work per system). The systems were: log4j-1.2.16, jabref-2.3, tvbrowser-2.7.6, 
rssowl-2.0.6, checkstyle-5.3, apache-commons (of Dec~2010), conqat-core-2.6, conqat-bundles-2.6, 
fckeditor-2.6, freemind-0.8.1, hibernate-3.6.0, openproj-1.4, sweethome3d-3.0, 
thight-vnc-java-1.3.10, and tomcat-6.0.24. For each of these systems, the 
expert-based measures were collected independently by two experts. The results 
were compared and differences in the expert judgement were discussed until a 
common judgement was achieved.


\subsection{Procedure}
The first step in our study is to collect the study data. For this purpose, we apply
the Quamoco quality evaluation approach to the study objects using the Quamoco
base quality model. As a result, for each study object, we get the measurement 
data for all measures of the quality model, as well as the aggregated values for 
the quality aspects. The full data is available in the PROMISE repository\footnote{\url{http://promisedata.googlecode.com}}. 
As we concentrate on the maintainability part of the base model, we only use
factors and measures related to maintainability.

In the next step, we create and evaluate the predictors for RQ~1. As independent 
variables all measures of the quality model are used, while as dependent 
variable, we use the evaluation of the quality aspect~\emph{maintainability} 
of the quality model. The experiments are implemented in a combination of Java 
and R\footnote{\url{http://www.r-project.org}}. R provides packages for all the 
predictors we use and it implements the statistics for evaluating the predictors. 
The Java program calls R~scripts and is used to implement the forward/backward 
selection and the cross-validation. We use a 4-fold cross validation, which means
the available study objects are partitioned into four subsets. Of the four subsets, one 
subset is used for validating the performance of the predictor, while the other 
three subsets are used for creating the predictor. This process is repeated four times,
with each subset used as validation data once. As the result for the MAR statistic
we use the average of the four runs. We also experimented with other folds (10-fold and
20-fold cross-validation), but the results did not change significantly.

For RQ~2, the predictors are built using all study objects of RQ~1. Since, for the
study objects of RQ~1 the expert-based measures are not available, these predictors
do not include the expert-based measures. These predictors are then evaluated by
predicting for the 15 study objects of RQ~2. 

\section{Results and Discussion}
\label{sec:results}

\begin{figure*}[p]
	\centering
	\includegraphics[width=.7\textwidth]{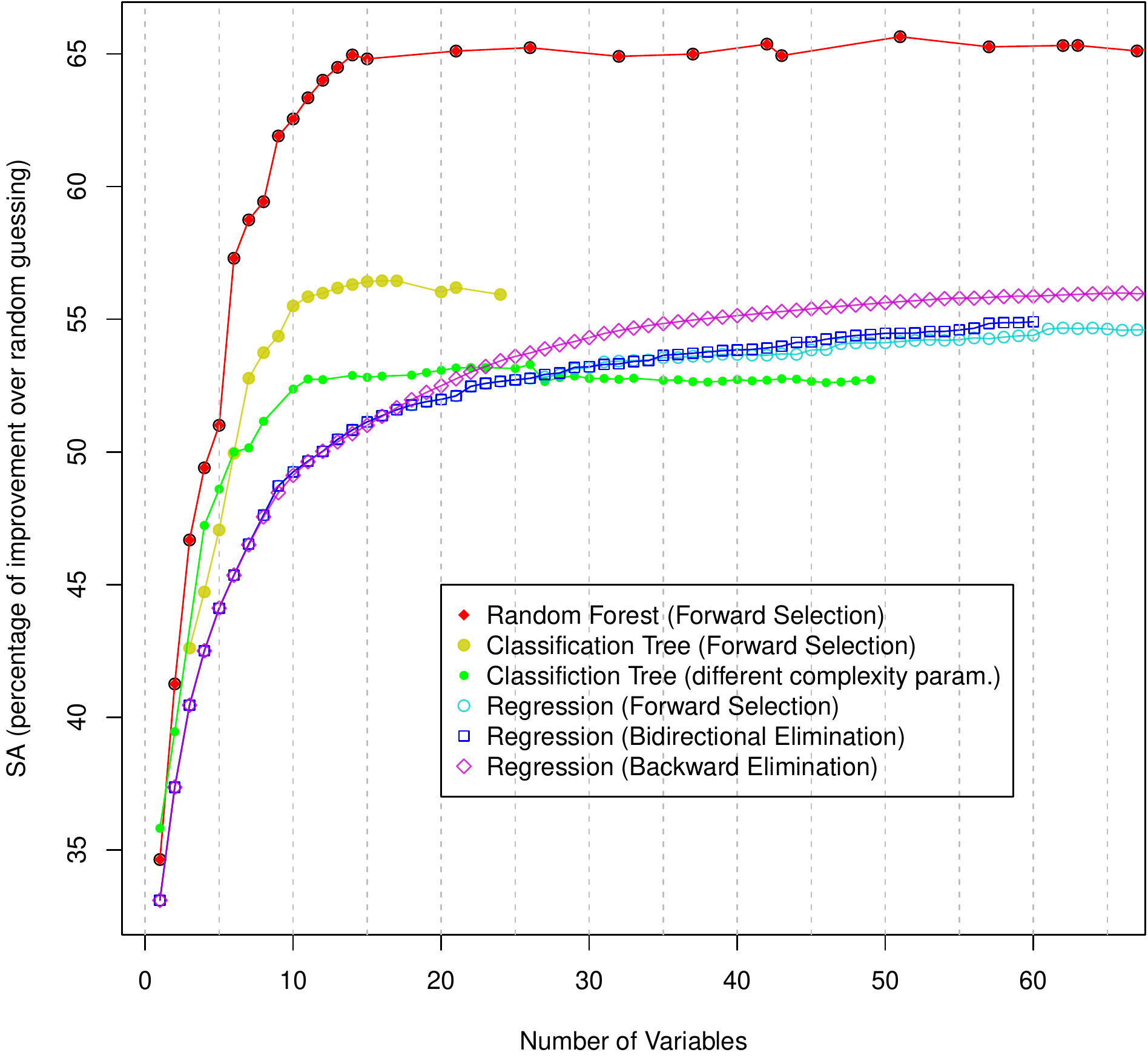}
	\caption{Performance of Predictors Depending on Numbers of Variables Used}
	\label{fig:sa:vs:variables}
\end{figure*} 

\begin{table*}
	\centering
	\begin{tabular}{|l|c|c|} \hline
		& \multicolumn{2}{c|}{Number of Variables} \\ \cline{2-3}
		Predictor 																& minimal & 10\,\%-best \\ 
							 																& MAR	& MAR \\ \hline
		Random Guessing  													& --& --\\
		$5\,\%$-qtl.\ of Random Guess. 							& --& --\\
		Linear Regr.~(Forw.~Sel.)									& 62 & 13 \\
		Linear Regr.~(Backw.~Elim.)								& 60 & 13\\
		Linear Regr.~(Bidir.~Elim.)								& 87 & 18 \\
		Class.\ Tree (dif.\ cmplx.\ parm.)				&	26 & 6 \\
		Class.\ Tree (Forw.~Sel.)									&	16 & 7 \\
		Random Forest (Forw.~Sel.)								& 51 & 10 \\ \hline
	\end{tabular}
	\caption{Optimal Number of Variables for the Predictors}
	\label{tab:optimal:number:of:variables}
\end{table*}

\begin{table*}
	\centering
	\begin{tabular}{|l|l|r|l|l|l|l|} \hline
		\multirow{2}{2cm}{Predictor} & \multicolumn{3}{c|}{minimal MAR} & \multicolumn{3}{c|}{10\,\%-best MAR} \\\cline{2-7}
																											& MAR 		& \multicolumn{1}{c|}{SA}	& ~$\Delta$ 	& MAR 	& \multicolumn{1}{c|}{SA}	& ~$\Delta$ 	\\ \hline
		Random Guessing  																	& 1.07 		& 0.00\,\%			&					 	& 			&					&						\\
		$5\,\%$-quantile of Random Guessing 								& 1.05		& 2.39\,\%			&						&				&					&						\\
		Linear Regression (Forward Selection)							& 0.49		& 54.67\,\%			&	0.61			& 0.53	&	50.47\,\% &	0.57 \\
		Linear Regression (Backward Elimination)					& 0.48 		& 54.90\,\% 		&	0.62			& 0.53	& 50.47\,\% & 0.57 \\
		Linear Regression (Bidirectional Elim.)						& 0.47 		& 56.16\,\% 		&	0.63			& 0.51	& 51.96\,\% & 0.58 \\
		Classification Tree (different complexity param.)	& 0.50 		& 53.29\,\% 		&	0.60			& 0.53  & 50.15\,\% & 0.56 \\
		Classification Tree (Forward Selection)						&	0.47		&	56.45\,\%			&	0.73			& 0.48  & 54.36\,\% & 0.57 \\
		Random Forest (Forward Selection)									& 0.37		& 65.64\,\%			&	0.74			& 0.40  & 60.90\,\% & 0.70 \\ \hline
	\end{tabular}
	\caption{Statistics for Predictors}
	\label{tab:statistics:for:predictors}
\end{table*}

In this section, we describe the results of the experiments and an interpretation 
according to the research questions of Section~\ref{sec:approach}. 

\subsection{RQ~1: Performance of Focused Quality Models}

As outlined in Section~\ref{sec:approach}, we built six predictors. We evaluated the 
performance of each predictor for varying numbers of variables used for prediction. 
Figure~\ref{fig:sa:vs:variables} shows the SA~statistics of the predictors depending on 
the number of variables used by the predictor. All predictors perform better with more 
variables although there are also slight degradations for, e.g.\ classification tree with 
forward selection. Most predictors have an SA between 30\,\% and 55\,\% with below 10 variables, 
which is already a reasonable improvement to random guessing. The maximum SA is about 65\,\% 
achieved by the random forest with forward selection. 

According to the approach of Section~\ref{sec:approach} to compare the predictors,
we determine the number of variables where each predictor performs best. Additionally,
we determine the minimum number of variables, for which the performance cannot be
improved by more than 10\,\% by adding more variables. Table~\ref{tab:optimal:number:of:variables} 
gives the details. The minimal number of variables for the optimal MAR ranges between 16 and 87. 
This seems to be due to slight changes in the MAR. If we look for a 10\,\% improvement,
the number of variables is reduced to 6--18. This conforms also to our observation that in Figure~\ref{fig:sa:vs:variables}
several of the predictors have achieved a close to maximum SA at about 10 variables.

\begin{table*}
	\centering
	\begin{tabular}{|l|l|l|l|l|l|l|} \hline
		Number of Variables & 6 & 8 & 10 & 12& 14 & 16	 \\ \hline
		Data set without expert-based measures		&  55.81\,\% & 59.76\,\% & 60.90\,\% & 63.25\,\% & 63.73\,\% & 63.90\,\%  \\
		Data set with expert-based measures				&  32.11\,\% & 33.75\,\% & 35.27\,\% & 43.16\,\% & 42.51\,\% & 46.72\,\%  \\
		\hline
	\end{tabular}
	\caption{SA statistic for data sets with/without expert-based measures}
	\label{tab:rq2:different:validations}
\end{table*}

Table~\ref{tab:statistics:for:predictors} gives the MAR, SA and $\Delta$ statistics
for the predictors with the number of variables determined above.

Regarding the performance of different predictors, in Figure~\ref{fig:sa:vs:variables} we 
see that the random forest predictor performs best with some distance to the other predictors. 
A reasonably good prediction is achieved with 10~variables with the MAR statistic only 
decreasing by 10\,\% over the best achievable MAR with 51~variables. The SA metric shows 
the random forest predictor with 10 variables is working 61\,\%~better 
than the baseline random guessing predictor. When using 51~variables it achieves an increase of 66\,\%
over the random guessing predictor. All other predictors achieve an SA value no better than 56\,\% which 
is considerably lower than the value of the random forest predictor. Overall, we see a reasonable
positive behaviour of the predictors.

The $\Delta$ statistic is about \mbox{$\approx0.6$} for all predictors, except for random forests and 
the classification tree with forward selection, which reach \mbox{$\approx0.75$}. 
Considering the smaller, 10\,\%-models, only the random forest still achieves 0.70.
This is a \emph{medium} to \emph{high} effect size. Therefore, the reasonable improvement over
random guessing has also a medium to high effect. Based on this, we conclude that a random forest 
predictor using 10 variables is well suited to be used as a focused quality model. 
Other predictors with roughly the same number of variables would still yield reasonable results.

\subsection{RQ~2: Performance Including Expert-Based Measures}

In the second research question, we used a random forest predictor, because it proved itself 
best in RQ~1. The random forest predictor was built using all study objects of RQ~1, and therefore
it does not include the expert-based measures. It was then used for predicting the quality 
assessment results for 15 systems, which included expert based measures.

\begin{figure}
	\centering
	\includegraphics[width=\columnwidth]{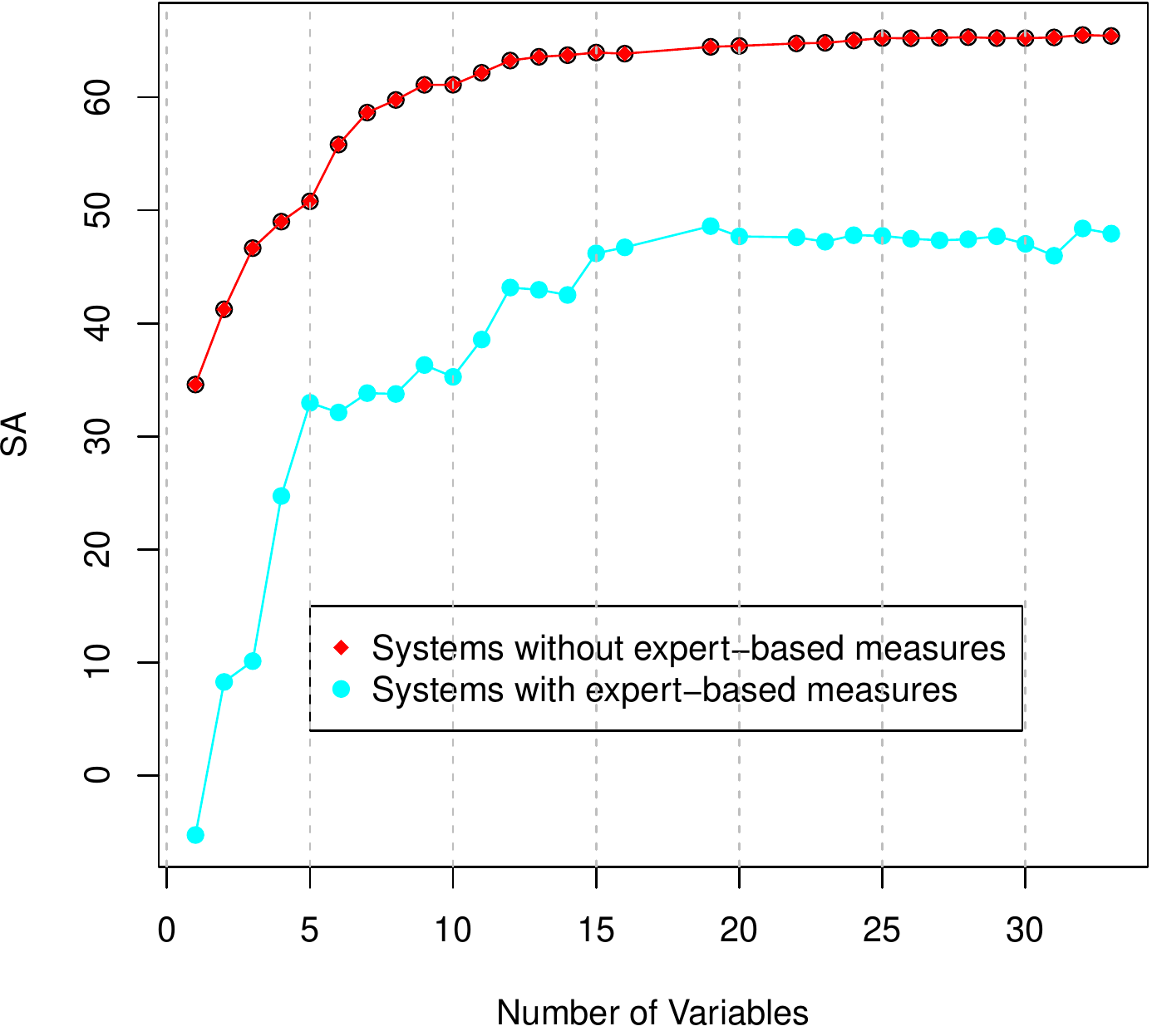}
	\caption{Random Forest Predictor used on data sets including and excluding expert-based measures}
	\label{fig:rq:2}
\end{figure}

Figure~\ref{fig:rq:2} and Table~\ref{tab:rq2:different:validations} show the SA statistic 
for different numbers of variables for two different predictors and data sets: (1)~the data 
without expert-based measures is the result of the cross-validation predictor of RQ~1 
(2)~the data with expert-based measures is the result of the predictor especially built 
for RQ~2 and applied to the 15 systems mentioned above. The SA statistic for the prediction
of the data including expert-based measures is considerably lower than in the results of RQ~1.
The SA statistic reaches its maximum of $48.60$ with a random forest using $17$ independent variables.
Thus, the prediction performance is worse than that of the worst predictor in RQ~1 (when
considering SA values for equal numbers of variables). Therefore, we conclude that the 
expert-based measures play an non-neglectible role. Accordingly, a predictor not considering 
these measures performs comparatively badly.

\section{Threats to Validity}
\label{sec:threats}

Our investigation faces several threats to the validity of our results. We discuss
internal and external validity threats as well as our mitigation strategies.

\subsection{Internal Validity}
One threat to internal validity is the choice and interpretation of the
MAR, SA, and $\Delta$ statistics. For minimising this threat, we aligned
our research according to published guidelines as described in the study 
design.

In RQ~2 a threat to validity is the low number~(15) of test systems. It
could distort the resulting values for the SA metric. For assessing this threat,
we conducted an additional experiment, to investigate the variance of the SA
statistic when only 15 systems are used. We repeated the experiments of RQ~1, but
calculated the SA statistic by predicting the values of only 15 randomly chosen 
systems (1000 tries). The standard deviation of the SA statistic was $0.0795$. 
Since, the values of SA were $0.6077$ and $0.4586$ respectively, we regard this
threat as minor.

A third threat to internal validity is the set of predictors we tested. Maybe
other prediction approaches (such as neural networks) could yield better results.
However, since we regard the prediction performance of random forests as satisfactory
this is no threat to our conclusions.

Finally, it is a threat to validity that in our experiments that showed correspondence
with expert opinion~\cite{Wagner.2012}, we included the manual measures but
in our experiments for RQ~1 we could not include them as they are not available
for so many systems. Hence, we cannot ensure that the Quamoco base model is
really valid in that case. Therefore, we included RQ~2 to specifically look at the influence of
manual measures.

\subsection{External Validity}
To generalise our results, the data and the systems we employed in our
study need to be representative for a broader population. We made sure that
our sample of open source systems contains a wide spectrum of systems
with different sizes and different types of applications. 

Despite the variations in the systems, they have all been built in Java. Several
of the measures we used are specific for this programming language. Hence,
the results may be different for other languages with other types of findings. 
As Java is a widely used programming language and in performing the study
we did not observe something that seemed specific to Java, we expect this
threat not to jeopardise the final conclusions.

We mostly employed automatic static analysis, because it is feasible for 
any company to collect data this way. Nevertheless, other types of analysis
could give different results. We consider our restriction as necessary to be able
to perform the study with this number of systems. Furthermore, it is comparable
to the large amount of existing work on models based on static analysis data.

Finally, we used a quality model focusing on the quality aspect of
maintainability. This could impair the generalisability to other
quality aspects. However, other studies (cf.~\cite{Menzies.2007,Neuhaus.2007}) 
show that predicting security and defects based on static code analysis 
works well and, hence, we assume the same holds for maintainability.


\section{Related Work}
\label{sec:related}

In the research area of defect prediction, we can find a lot of work on source code metrics
and their statistical analysis. Fenton~et~al.~\cite{Fenton.1999} and
Hall~et.~al.~\cite{Hall.2011} report on literature surveys summarising a large 
number of publications on defect prediction. Most of these studies have a different 
focus than our study. They build predictors for predicting faulty components within 
one system. We instead, build predictors being applied to a large number of software 
systems and predicting a given quality evaluation result instead of faultiness of
components. Furthermore, these studies use general code metrics like complexity or
lines-of-code as independent variables, while we use hundereds of static-code analysis 
rules. Nevertheless, some conclusions of these studies are relevant for our study. 

Hall~et.~al.\ concludes that no general recommendation for a certain prediction 
technique can be given. Rather, depending on diverse context factors, different 
prediction techniques work better or worse. This supports our approach of using 
several prediction techniques and comparing the results. 


To the best of our knowledge there is no study directly resembling our approach
of using predictors to predict a quality evaluation result from a comprehensive
quality model. The approach
of Nagappan and Ball~\cite{Nagappan.2005} comes closest to our approach, because
they use the rules of static code checkers as independent variables, like we do. 
However, they again try to predict faulty components based on it with quite 
successful results. We take this result as an indicator that rule-based static 
code analysis is principally suited for quality analysis.

Another similar approach to ours is that of Wagner~\cite{Wagner.2010c}. He described 
an approach to express a quality model similar to the Quamoco model using Bayesian 
nets. He found that the structure of the quality model transforms well into a Bayesian 
net but the prediction results were mixed. A reduction of measures was not considered.

While Hall~et.~al.\ generally do not recommend a certain prediction technique, 
there is some evidence, that in certain contexts random forests work best. For
the NASA data set, both Guo~et~al.~\cite{Guo.2004} and Lessmann~et.~al.~\cite{Lessmann.2008}
show a better performance of random forests over other prediction techniques. 
This result is in line with out results, where the random forest also worked best. 


\section{Conclusions and Future Work}
\label{sec:conclusions}

For evaluating the quality of software systems, comprehensive quality models
have been built to reflect the diversity in software quality factors. The models 
rely on large numbers of measures (usually some hundreds) 
to get to a quality evaluation. On the other hand, there are focused quality models
making use of a small set of measures for evaluating quality as well. Such a focused
quality model has advantages as less effort is necessary for collecting data.
If the inclusion of a high number of measures in comprehensive quality models really
gives significantly different results than focused quality models has not been
investigated so far.

Our results from comparing the Quamoco base model to focused models built
using machine learning show that it is possible to build an evaluation model 
only relying on 10~measures while having a standardised accuracy of~61\,\%. 
The reduction from 378~measures in the comprehensive model to 10~measures facilitates 
the application in practice, because only 10~measures need to be calculated and, e.g.,
checked for false positives. However, if the evaluation model is to be used as a
basis for contracting or design guidelines, it must be kept in mind that the results
produced by the reduced model may differ from the results produced by the comprehensive 
model. 

A major advantage of the focused model would be, if its reduced number of
measures enabled to omit expert-based measurements. Our results indicate
that the expert-based measures play a role that should not be neglected. The evaluation
model built without expert-based measures works considerably worse for 
predicting assessments which include expert-based measures.

We put specific emphasis on the repeatability of our results. The used quality model
and tool support for quality evaluation is freely available on the Quamoco web site\footnote{\url{http://www.quamoco.de/}}.
We further publish our data and results in the PROMISE\footnote{\url{http://promisedata.googlecode.com}}
repository so that future research can easily reassess and improve our findings.

For future work we plan to investigate different constraints for selecting
the measures used in the predictor. Possible constraints could be the number 
of different tools needed to calculate the measures, the price and effort 
for using a tool or the preciseness of measures.

\section{Acknowledgments}
The presented work was partially funded by the German Federal Ministry of Education and Research~(BMBF), grant ``Quamoco, 01IS08023B''.

\bibliographystyle{abbrv}
\bibliography{light-weight-qm}

\end{document}